\documentclass[prd,preprint]{revtex4}
\usepackage{tabularx}
\usepackage[dvips]{color}
\usepackage[open]{bookmark}
\renewcommand\L{\mathcal{L}}

\newcommand{\AER}[1]{{\color{black} #1}}

	\usepackage{amsmath}
	\usepackage{amsfonts}
  	\usepackage{amssymb}
  	\usepackage{float}
	\usepackage{makeidx}
	\usepackage{amsfonts}
	\usepackage[ansinew]{}
	\usepackage[usenames,dvipsnames]{pstricks}
	\usepackage{epsfig}

\usepackage{xcolor}

\newcommand{\be}{\begin{equation}}
\newcommand{\ee}{\end{equation}}
\newcommand{\bea}{\begin{eqnarray}}
\newcommand{\eea}{\end{eqnarray}}

\makeindex

\begin{document}

\preprint{CERN-TH-2023-013}

\title{Gravitoelectromagnetic quadrirefringence}
\author{Antonio Enea Romano $^{1,2}$, Sergio A. Vallejo-Pe\~na$^{1,2}$}

\affiliation{
${}^{1}$
Instituto de F\'isica, Universidad de Antioquia, A.A.1226, Medell\'in, Colombia \\
${}^{2}$ICRANet, Piazza della Repubblica 10, I--65122 Pescara, Italy}



\begin{abstract}
We develop an effective approach for the study of the interaction of gravitational waves (GWs) and electromagnetic waves (EMWs), showing that   quadrirefringence can be produced, a phenomenon consisting in a frequency and polarization dependency of the speed of the different polarizations of GWs and EMWs, \AER{which is also inducing a frequency and polarization dependent modification of the GW-EMW luminosity distance ratio.}
Quadrirefringence can be due to the GW-EMW interaction in the source or during the propagation from the source to the observer. 
In the first case the astrophysical properties of the source can induce a unique characteristic imprint on GWs and EMWs for each source, the  rainbow of binaries, while the effect on the propagation from the source could be used to probe the large scale electromagnetic field using GWs.
Numerical general relativistic magnetohydrodynamic (GRMHD) simulations can be used to obtain precise theoretical predictions for different binary systems with large electromagnetic and gravitational  fields, such as  binary neutron stars.

\end{abstract}

\pacs{Valid PACS appear here}
\maketitle



\section{Introduction}
The detection of gravitational waves (GWs) \cite{LIGOScientific:2016aoc} by the Laser Interferometer Gravitational Wave Observatory (LIGO) and Virgo has started the era of multi-messenger astronomy. Electromagnetic counterparts \cite{LIGOScientific:2017zic,LIGOScientific:2017vwq,Goldstein:2017mmi,Coulter:2017wya,Coughlin:2018fis} are particularly useful to test the general relativity prediction that GWs and EMWs propagate in the vacuum at the same speed \cite{Baker:2017hug,Creminelli:2017sry,Sakstein:2017xjx,Ezquiaga:2017ekz,Wang:2017rpx}, and the interaction between GWs and EMWs in the source or  during the propagation from the source to the observer. 

\AER{The phenomenon of gravitational waves production by the electromagnetic waves, also known as Gertsenshtein effect, \cite{Gertsenshtein,Palessandro:2023tee,Johnston:1974vf}, is nornally is computed ignoring the back-reaction of GWs on EMWs, and vice versa the gravitationally induced electromagnetic radiation was studied without including the back-reaction of EMWs on GWs
 \cite{Johnston:1973cd}.
In this paper we go beyond these approximations, and derive the equations accounting for the full interaction between GWs and EMWs, including the back-reaction of GWs on EMWs, and of EMWs on GWs.  
We then develop a formalism to compute the effective speed of GWs and EMWs and show that both types of waves are expected to experience birefringence, a phenomenon we call  quadrirefringence, which naturally arises as a direct implication of the covariant formulation of electromagnetism in curved space. }
Quadrirefringence modifies the propagation of different polarizations of the GWs and EMWs, making the effective propagation speed of each polarization, time and frequency dependent.
Quadrirefringence can be understood as the consequence of the fact that for EMWs, GWs act as an effective medium, and viceversa for GWs, EMWs  act as an effective medium as well. 
When quadrirefringence happens in the source, it leaves a characteristic imprint on GWs and EMWs, which can be different for each binary system leading to the rainbow of binaries.
On cosmological scales quadrirefringence can instead be used to probe the large scale electromagnetic field.
In this paper we establish a general theoretical framework to study the GW-EMW interaction, and leave to future works the calculation of the physical effects for realistic astrophysical or cosmological scenarios.

\section{Gravitoelectromagnetism Lagrangian}

The minimal coupling of gravity to electromagnetism is given by the Lagrangian density \cite{inverno:1992,weinberg2008cosmology,Cote:2019kbg}
\bea
\L&=&\L_G+\L_{EM}=\sqrt{-g}\Big[R+\frac{1}{4}F_{\mu\nu}F^{\mu\nu}\Big]=\sqrt{-g}\Big[R+\frac{1}{4}g_{\mu\nu}g_{\rho\sigma}F^{\mu\rho}F^{\nu\sigma}\Big] \,,\label{LEM}
\eea
where the Faraday's tensor is defined as $F_{\mu\nu}=\partial_\mu A_{\nu}-\partial_\nu A_{\mu}$ in terms of the vector potential $A^{\mu}$, $R$ is the Ricci scalar, $g$ is the determinant of the metric, and we are using units in which $8\pi G=c=\epsilon_0=1$, where $G,c,\epsilon_0$ are the Newton's gravitational constant, the vacuum speed of light and the vacuum permittivity.
In the following we will treat both GWs and EMWs as perturbations.

For a gravitational wave propagating along the $x$-axis, the leading order interaction Lagrangian density in the Einstein's gauge is given by the cubic terms
\be
\L_{EM}=\frac{1}{2} (F^{02}F^{02}-F^{03}F^{03}-F^{12}F^{12}+F^{13}F^{13}) h_+ + (F^{02}F^{03}-F^{12}F^{13})h_{\times}=\Pi_+ h_+ +\Pi_{\times} h_{\times} \label{LEMP}\, ,
\ee
 where the perturbed metric relevant for the computation of the effects of the graviton photon interaction is
\be
g_{\mu \nu}=\begin{pmatrix}
1 & 0 & 0 & 0 \\
0 & -1 & 0 & 0 \\
0 & 0 & -1+h_+ & h_{\times} \\
0 & 0 & h_{\times} & -1-h_+ \\
\end{pmatrix} 
\, .
\ee

From eq.(\ref{LEMP}) it is evident that the  different polarizations of the GWs and EMWs are coupled in different ways to each other, which is the origin of the quadrirefringence.
\section{GWs propagation equation}
In the Einstein's gauge the Lagrangian density for GWs, including the leading order minimal coupling to electromagnetism, is 
\be
\L_h=h_A'^2-(\partial h_A)^2+\L_{EM} \,,
\ee
which gives the equations of motion
\bea
h_A''-\nabla^2 h_A&=\Pi_A \, , \label{GWseq}
\eea
where $\Pi_A$ was given in eq.(\ref{LEMP}), \AER{and the sub-index $A$ stands for the two polarizations  $+$ and $\times$.}
Not that the  $\Pi_A$ are different for each polarization mode, due to the structure of the electromagnetic stress-energy-momentum tensor. Similar results are expected when considering other gauge fields, such as the axion.
In the Lagrangian approach $\Pi_A$ is simply arising from the coupling $\Pi_A h_A$, given explicitly in eq.(\ref{LEMP}).

\section{EMWs propagation equation}
The equations of motion for the EMWs can be derived by varying the action with respect to $A^{\mu}$, giving the covariant form of the Maxwell's equations

\be
\nabla_{\mu}F^{\mu\nu}=\frac{1}{\sqrt{-g}}\partial_\mu(\sqrt{-g}F^{\mu\nu})=0 \, ,
\ee
where the determinant of the perturbed metric is 
\be
\sqrt{-g}=(1-h^2_+-h^2_{\times})^{1/2}\approx 1-\frac{1}{2}(h^2_++h^2_{\times})\,.
\ee
This metric determinant is inducing the graviton photon coupling in the Maxwell's equations, which take the form
\be
\partial_{\mu}F^{\mu\nu}=J_{eff}^{\nu}=-\frac{\partial_{\mu}\sqrt{-g}}{\sqrt{-g}}F^{\mu\nu} \,,
\ee
where we are denoting  the effective gravitoelectromagnetic current as $J_{eff}^{\nu}$. The gravitoelectromagnetic current  arises naturally as a consequence of the photon graviton interaction. 

Imposing the Lorentz's gauge condition $\partial_{\mu}A^{\mu}=0$, and taking the leading order interaction terms in the determinant of the metric, we get the wave equation   
\be
\Box A^{\nu}=J_{eff}^{\nu} = (h_+\partial_{\mu}h_++h_{\times}\partial_{\mu}h_{\times})F^{\mu\nu} \label{emeq} \,.
\ee
\section{Quadrirefringence}
The system of coupled differential equations
\bea
\Box h_A&=&\Pi_A(A^{\rho}) \, , \\
\Box A^{\nu}&=&J_{eff}^{\nu}(h,A^{\rho}) \, ,
\eea
describes the evolution and interaction of GWs and EMWs, and their interaction is associated to the source terms which couple the two equations.
In the above equations we have used an abstract notation to make explicit the dependence on $h$ and $A^{\rho}$ of the source terms given in eq.(\ref{GWseq}) and eq.(\ref{emeq}), to underline the coupling between the two equations. Quadrirefringence is due to the fact that the source can be different for different polarizations of the GWs or different components of $A^{\nu}$.
The effects of the polarization of EMWs can be obtained by expressing the electric and magnetic fields in terms of $A^{\nu}$ using
\bea
\bf{E}=-\nabla{\phi}-\partial_t \bf{A} \, , \\ 
\bf{B} = \nabla \times \bf{A} \, , 
\eea
where $A^{\nu}=(\phi, \bf{A})$.

Since the GWs propagating in the $x$ direction have the form $h(\eta-x)$,  we get
\be
J^{\nu}_{eff}=(h_+h'_++h_{\times}h'_{\times})F^{0\nu}+(h_+\partial_{x}h_++h_{\times}\partial_{x}h_{\times})F^{1\nu}\,,
\ee
\AER{where $\eta$ is the conformal time.}
Note that each component of the effective current is different from zero, implying that each component of $A^{\mu}$ is affected by quadrirefringence.

If the effects of interaction are ignored the EMWs and GWs will propagate independently, and the solutions of the wave equations will be of the standard oscillatory form, with each polarization mode of the EMWs and GWs  propagating at the speed of light, but if the effects of interaction are taken into account, each polarization of the GWs and EMWs will propagate differently, since the source terms are different.

The strength of these effects depends on the size of the sources, i.e. $\Pi_A$ and $J^{\nu}_{eff}$, which for GWs is related to the anisotropic part of the stress-energy-momentum tensor of the electromagnetic field, and for EMWs to the amplitude of the GWs.

\section{The rainbow of bright and grey sirens}
So far we have not introduced any electromagnetic four current $J^{\nu}_{EM}$ in the Lagrangians, i.e. we have studied the interaction of GWs and EMWs in the absence of matter.

In a realistic astrophysical scenario $J^{\nu}_{EM}$ will be definitely present in the source, and will be associated to the specific dynamic of the astrophysical system producing the EMWs and GWs, which could be a neutron stars (bright siren) or neutron star black hole (grey siren) binary.

In this case the additional Lagrangian density $\L_{mat}$ associated to the interaction with matter, and corresponding equations of motion would take the form
\bea
\L_{mat}&=&\sqrt{-g}A_{\mu}J^{\mu}_{EM}\,, \\
\Box h_A&=&\Pi_A(A^{\rho}) \,, \label{EQGWs} \\ 
\Box A^{\nu}&=&J_{eff}^{\nu}\,(h,A^{\rho})+J_{EM}^{\nu}=J_{tot}^{\nu}\,. \label{EQEMWs}
\eea
Due to the coupling between the  equations, the current $J_{EM}^{\nu}$ will also affect the GWs emitted by the the binary. 
Given the environmental dependence of $J^{\nu}_{EM}(\eta,x^i)$ and $\Pi_A(\eta,x^i)$, i.e. the fact that its space  and time evolution depend on the specific dynamics of each astrophysical system, the effects of quadrirefringence can be different for each binary system, a phenomenon we dub "binary rainbow".  

Since the interaction effects depend on the specific properties of each binary system, fully general relativistic magnetohydrodynamic (GRMHD) simulations \cite{Cipolletta:2019geh,Etienne:2015cea}  are necessary to compute them for each binary. This a complicated numerical problem which goes beyond the scope of this paper, which is to establish a theoretical framework for the study of GW-EMW interaction, but is an important future task to check the observability of these effects.
If the predicted effects are within the sensitivity of present or next generation GWs and EMWs observatories, multimessenger observations will allow to infer new information about the binary system dynamics, by performing a combined analysis of GWs and EMWs observations, in particular of their polarizations.

\section{Effective speed approach}
It is possible to describe quadrirefringence using an effective  approach \cite{Creminelli:2014wna,Gleyzes:2013ooa,Romano:2023uwf,Gubitosi:2012hu}, in which the effects of interaction are encoded in an appropriately defined polarization, frequency, and time dependent effective speed of GWs and EMWs.
While quadrirefringence can be studied without adopting this effective description, by solving the system of coupled differential equations in eqs.(\ref{EQGWs}-\ref{EQEMWs}), it is convenient to give this equivalent effective description, to gain more insight about the physical implications.

Let's denote as $\hat{A_{\nu}}$ a solution of the system of coupled differential equations given in eqs.(\ref{EQGWs}-\ref{EQEMWs}), and with $\hat{J}_{\nu}$ the current obtained substituting the same solutions. We have lowered the indices to easy of notation. We can manipulate the equation for EMWs as following
\bea
\hat{A}''_{\nu}-\nabla^2 \hat{A}_{\nu}-\hat{J}_{\nu}&=& (\hat{A}'_{\nu})'-\left(\int \hat{J}_{\nu}d\eta\right )'-\nabla^2 \hat{A}_{\nu}=\left[\hat{A}'\left(1-\frac{\hat{g}_{\nu}}{\hat{A}'_{\nu}}\right)\right]'-\nabla^2 \hat{A}_{\nu}=0\,,
\eea
where we have defined $\hat{g}_{\nu}=\int \hat{J}_{\nu} d\eta$.
After defining the effective speed as 
\be
c_{\gamma,\nu}^2=\left(1-\frac{\hat{g}_{\nu}}{\hat{A}'_{\nu}}\right)^{-1} \label{cEMeff}\,,
\ee
the equation can be written as
\be
\hat{A}''_{\nu}-2\frac{c'_{\gamma,\nu}}{c_{\gamma,\nu}}\hat{A}'_{\nu}-c_{\gamma,\nu}^2 \nabla^2 \hat{A}_{\nu}=0 \label{EMeff}\,.
\ee

This proofs that a solution of eqs.(\ref{EQGWs}-\ref{EQEMWs}) is also a solution of eq.(\ref{EMeff}), for an appropriate definition of the effective speed, as given in eq.(\ref{cEMeff}). 
The effective speed  $c_{\gamma,\nu}(\eta,x^i)$ is  space and polarization dependent, since it is obtained from space and polarization dependent quantities, which in momentum space lead to a frequency dependency, and is  related conceptually to the dispersion relation in plasma physics , but in our case the role of medium for EMWs is played by gravitational waves.
The calculation of the effective speed requires the substitution of the solutions of the system of differential equations, so unless the equations have simple analytical solutions, they have to be obtained numerically, and different initial conditions will give different effective speeds \cite{Romano:2020oov}.

In this approach the effects of the current on the propagation of the EMWs has been encoded in the effective speed. Since the current can be  polarization, time and space dependent, the effective speed can also have the same dependency. This is in agreement with the experimental evidence of the change of the speed of EMWs, depending on the the medium in which are propagating. In this approach the physical properties of the medium are modelled mathematically by the effective speed.
This definition of effective speed is model independent, and could be applied to obtain an effective description of the propagation in any medium, and in this paper we consider the effects of the total current given by the sum of the EM current and the effective gravitomagnetic current.

The dumping terms in eq.(\ref{EMeff}) is related to the change in the amplitude of the EMWs wave due to its interaction with  gravitational waves, which act effectively as an isotropic medium.

In a similar way \cite{Romano:2022jeh,Romano:2023uwf}, an effective equation can be derived for GWs, 

\be
\hat{h}_A''-2\frac{c'_{T,A}}{c_{T,A}}\hat{h}'_A-c_{T,A}^2 \nabla^2 \hat{h}_A=0 \, ,\\
\ee
where $c_{T,A}(\eta,x^i)$ is the GWs space dependent effective speed (SES), defined as
\bea
\hat{g}_A&=&\int \hat{\Pi}_A d\eta \,,\\
c^2_{T,A}&=&\left(1-\frac{\hat{g}_A}{\hat{h}_A'}\right)^{-1} \,, \label{cGWseff}
\eea
and  again the hat denotes solutions, or quantities obtained by substituting the solutions of the system of differential equations. \AER{ Note that in the effective equation (\ref{cGWseff}) there is a damping term, which is related to the effects of the interaction on the GW amplitude.

In \cite{Romano:2022jeh,Romano:2024apw} it was shown that in general the friction term  induces a frequency and polarization dependent GW-EMW luminosity distance ratio, given by the ratio of the effective speed at the source and the observer, according to
\be
\frac{d_{L}^{GW}}{d_{L}^{EM}}(z)=\frac{\tilde{c}_{T,A}(z,k)}{\tilde{c}_{T,A}(0,k)}\,.\label{dgw}
\ee
This implies that another expected observable effect of quadrirefringence is  the modification of the GW-EMW luminosity distance ratio} 

It is important to observe that different initial conditions lead to different effective speeds, and that  while by construction the effective equations admit as solutions the solutions of the system of differential equations, they also admit other solutions, such as  for example $h_A=A_{\nu}=0$, or other solutions with initial conditions different from the ones used to solve the system of differential equations.
This is expected, since the effective equation is obtained by imposing that it admits one given solution, but this does not prevent other possible solutions.
The other solutions are not physical relevant for the purpose of the effective approach.

\section{Momentum  dependent effective speed}
In Fourier space a similar set of equations can be derived in terms of the  momentum dependent effective speed (MES)

\bea
\tilde{h}_A''-2\frac{\tilde{c}'_{T,A}}{\tilde{c}_{T,A}}\tilde{h}'_A+\tilde{c}_{T,A}^2 k^2 \tilde{h}_A=0 \, , \\
\tilde{A}''_{\nu}-2\frac{\tilde{c}'_{\gamma,\nu}}{\tilde{c}_{\gamma,\nu}}\tilde{A}_{\nu}+k^2 \tilde{c}_{\gamma,\nu}^2 \tilde{A}_{\nu}=0 \, ,
\eea
where a tilde denotes the Fourier modes, except for the MES $\tilde{c}_{T,A}(\eta,k)$ and $\tilde{c}_{\gamma,\nu}(\eta,k)$, which are not defined \cite{Romano:2022jeh,Romano:2018frb,Romano:2020oov,Romano:2023uwf} as the Fourier transform of the SES, but as

\bea
\tilde{g}_A&=&\int \tilde{\Pi}_A d\eta \,,\\
\tilde{c}_{T,A}^2&=&\left(1-\frac{\tilde{g}_A}{\tilde{h}_A'}\right)^{-1} \, \label{cGWseffk}
\eea

and
\bea
\tilde{g}^{\nu}&=&\int \tilde{J}_{\nu} d\eta \,, \\
\tilde{c}_{\gamma,\nu}^2&=&\left(1-\frac{\tilde{g}_{\nu}}{\tilde{A}'_{\nu}}\right)^{-1} \,. \label{cEMeffk}
\eea
In the above equations, for easy of notation, we have dropped the hat used previously to denote quantities obtained by substituting the solutions.

Note that the GW-EMW interaction not only changes the effective speed of each polarization, but it also induces an additional time-frequency-polarization dependent friction term, related to the time derivative of the MES, which can have important observational implications \cite{Romano:2023ozy,Romano:2022jeh} for the gravitational and electromagnetic luminosity distances, which we will study in a future work.

\section{Conclusions}
We have shown that the interaction between GWs and EMWs  induce the phenomenon of quadrirefringence, the different propagation of the polarization modes of GWs and EMWs.
When this phenomenon happens in binary systems like bright or grey sirens, it can induce an object characteristic imprint on the emitted GWs and EMWs, the binaries rainbow, which could provide a new method to study the astrophysical properties of these systems, such as for example the equation of state of neutron stars.
On cosmological scales quadrirefringence can be used to probe the electromagnetic field with gravitational waves. 

We have studied the leading order effect of the classical calculation of the GW-EMW interaction, but it will also be important to consider quantum effects, and to study the regimes in which they can provide important corrections to the classical effect.

Note that while in this paper we have focused on the GW-EMW interaction effects, similar effects are predicted by the model independent effective approach \cite{Romano:2022jeh}, and could arise also when the graviton is coupled to any other field.
These interaction effects will allow to use gravitational waves observations to probe any other field the graviton is coupled to.

It order test the observability of the GW-EMW interaction effects it will be important to perform numerical general relativistic magnetohydrodynamic (GRMHD) simulations to obtain precise theoretical predictions, and check the validity of the effective speed approach.


\begin{acknowledgments}
We thank Luca Baiotti, Sudipta Hensh, Filippo Vernizzi and Mairi Sakellariadou for interesting discussions. This work was supported by the UDEA projects  2021-44670,  2019-28270, 2023-63330.
\end{acknowledgments}


\bibliographystyle{h-physrev4}

\end{document}